\documentclass[aps,prl,twocolumn,superscriptaddress]{revtex4}

\usepackage{graphicx}

\begin{document}


\title{The mid-infrared AC Hall effect in optimally-doped Bi$_2$Sr$_2$CaCu$_2$O$_{8+\delta}$}

\author{D. C. Schmadel}
\email[]{schmadel@physics.umd.edu}
\affiliation{Department of Physics, University of Maryland, College Park, Maryland 20742 USA}

\author{J. J. Tu}
\affiliation{Department of Physics, Brookhaven National Laboratory, Upton, New York 11973 USA}

\author{D. B. Romero}
\affiliation{Laboratory for Physical Sciences, University of Maryland, College Park, Maryland 20742 USA}

\author{L. Rigal}
\affiliation{Department of Physics, University of Maryland, College Park, Maryland 20742 USA}
\affiliation{Laboratoire National des Champs Magn\'{e}tiques Puls\'{e}s, Toulouse, France}

\author{M. Grayson}
\affiliation{Department of Physics, University of Maryland, College Park, Maryland 20742 USA}
\affiliation{Walter Schottky Institut, Technische Universit\"{a}t M\"{u}nchen, D-85748 Garching, Germany}

\author{G. D. Gu}
\affiliation{Department of Physics, Brookhaven National Laboratory, Upton, New York 11973 USA}

\author{H. D. Drew}
\affiliation{Department of Physics, University of Maryland, College Park, Maryland 20742 USA}
\affiliation{Center of Superconductivity Research, University of Maryland, College Park, Maryland 20742, USA}


\date{November 21, 2002}

\begin{abstract}
A novel heterodyne polarometry system determines the frequency dependence from 900 to 1100 cm$^-1$ and temperature dependence from 35 to 330 K of the Hall transport in single crystal, optimally doped Bi$_2$Sr$_2$CaCu$_2$O$_{8+\delta}$. The results show a significant disconnect from the behavior of the Hall angle in the existing data for YBCO in the far-infrared, which indicate a negative value for the real part of the Hall angle above 250 cm$^{-1}$, whereas that of the current work at 1000 cm$^{-1}$ is positive.  The current work when analyzed using an extended Drude formalism results in a Hall mass comparable to the ARPES Fermi mass and a scattering rate comparable to the low frequency conductivity and Hall effect scattering rates, which, however, are only 1/4 of the minimum ARPES values.
\end{abstract}
\maketitle
While the DC and infrared conductivities reveal the overall response of a system to an electric field, the Hall effect reveals the sign of the carriers as well as some information as to the path taken by them.  It essentially probes the ability of the carriers to veer or bend their course.  Quantities frequently used to represent the effect are the Hall angle $\theta_{\text{H}}$ or the longitudinal and transverse conductivities $\sigma_{xx}$ and $\sigma_{xy}$ with
\begin{equation}
\tan \theta_{ \text{H}}=\frac{\sigma_{xy}}{\sigma_{xx}}
\end{equation}
These quantites become complex response functions for AC fields.  The collection of existing measurements of the DC and AC Hall effect in high temperature superconductors has revealed a most curious behavior, whose description defies all current models, while, however, apparently falling short of suggesting a comprehensive theortical approach on its own.  Some of the difficulty of identifying patterns which can be simply expressed by means of some phenmenology arises from the fact that one is hesitant to make direct comparisons between the the existing AC Hall measurements conducted on epitaxial films of YBCO with thoses of results of ARPES and STM on single crystal BSCCO.  The YBCO films are twinned on a micron scale and the conductivity is further complicated by the existance of CuO$_2$ chains.  The mid-infrared Hall measurments reported here are the first on BSCCO single crystal and should cover some of the distance in remedying this predicament.

To appreciate the present state of the search for patterns and phenomenology consider the first of the earlier work \cite{Ong1987, Chien1991, Harris1992}, which to an extent indicated that $\cot \theta_H$ followed squared temperature behavior:
\begin{equation}
\cot \theta_{ \text{H}}=A+BT^2
\end{equation}
with $A$ proportional to Zn doping.  The T$^{n}$ behavior, for n=2, supported two $\tau$ models such as those associated with spin-charge separation in the cuprates~\cite{Anderson1991}.  However, later systematic studies of both single layer Bi$_2$Sr$_{1.6}$La$_{0.4}$CuO$_y$ and bilayer Bi$_2$Sr$_2$CaCu$_y$~\cite{Konstantinovic2000} have revealed that n = 1.78 for optimal doping and n  clearly depends on oxygen doping, approaching n = 2 only in the underdoped regime.  The far-infrared extension of the Hall effect explored in YBCO~\cite{Kaplan1996, GraysonPRL2002} suggests a different phenomenology involving a squared Lorentzian with a scattering rate linearly proportional to temperature~\cite{GraysonTrieste2002}.  A theoretical justification for this response has been obtained by Varma and Abrahams in terms of marginal Fermi liquid model in the form ~\cite{Varma2001}:
\begin{equation}
\tan \theta_{\text{H}}=\frac{\omega_{\text{H}}}{\gamma_{\text{H}}}+\frac{\Omega \omega_{\text{H}}}{\gamma_{\text{H}}^2}
\end{equation}
where $\Omega$ is related to Fermi surface averages of derivatives of the elastic scatteering and $\gamma_{\text{H}}$ is the inelastic relaxation rate proportional to temperature and which is replaced with $\gamma_{\text{H}} - i\omega$ for ac.  In the dc limit this form accommodates the revised exponent of 1.78 and the Zn doping.  This far-infrared behavior contrasts sharply with that suggested by the mid-infrared data for YBCO~\cite{CernePRL2000} obtained using the spectral lines of a CO$_2$ laser.  When the cotangent of the Hall angle is fitted to a Drude model the resulting Hall mass and the Hall scattering rate, to within the accuracy of the data, display no frequency dependence within the range from 949 to 1079 cm$^1$.  However, the Hall frequency displays a slight decrease with increasing temperature, whereas the Hall scattering rate increases linearly with temperature from near zero at T$_{\text{c}}$ to over 300 cm$^-1$ at 250 K.  However, the chain contributions to the conductivity for these twinned samples, grown by pulsed laser deposition, introduces considerable uncertainty in these results for the scattering rates and Hall mass.

In developing an understanding of transport in the cuprates the relation between the transport properties and the Angle-Resolved Photoemission Spectra (ARPES) will be important. In the past several years ARPES has mapped the Fermi surface and identified the d-wave symmetry of the order parameter.  Using $k_{\text{F}}=0.446 \,\text{\AA}^{-1}$ and $v_{\text{F}} = 1.8 \,\text{eV\AA}$~\cite{VallaPRL2000, VallaScience1999} the ARPES Fermi surface mass is
\begin{equation}
m_{\text{F}}=\frac{\hbar k_{\text{H}}^*}{v_{\text{F}}}=\hbar k_{\text{H}}^* \left( \frac{1}{\hbar}\frac{\partial E}{\partial k} \right)^{-1}
\end{equation}
where $k_{\text{F}}^*= 0.69 \, \text{\AA}$ is the Fermi wavenumber for the hole-like surface.  The quasiparticle relaxation rate $\Gamma$ is related to the width of the ARPES spectral density function at the Fermi energy.  Within many body perturbation theory the low frequency current relaxtion rate is given directly by the EDC line widths or from the MDC line widths, $\Delta k$, as $\Gamma_{\text{F}} \approx v_{\text{F}} \Delta k$, where $v_{\text{F}}$ is the measured Fermi velocity.   The existing ARPES data on BSCCO finds that $\Gamma \propto \text{T}$.  However, the minimum values for $\Gamma$ deduced from ARPES corresponding to the $(\pi, \pi)$ or nodal direction on the Fermi surface are four times larger than the measured IR relaxation rates.  It is not yet known whether this discrepancy can be reconciled in terms of small angle scattering issues or Fermi liquid effects or if it requires a more exotic interpretation.

Additionally ARPES measurements have revealed that there is a kink in the quasiparticle dispersion curve, which for Bi$_2$Sr$_2$CaCu$_2$O$_{8+\delta}$ occurs at about 50$\pm$15 meV~\cite{Bogdanov2000}. The change in slope results in a reduction in velocity of around a factor of two above the kink.  Z. X. Shen \textit{et al.}~\cite{ShenCond-mat2001} also report this kink occurring in LSCO at around 70 meV.  The interprertation of this feature is controversial and includes electron-phonon interaction, the opening of a 50 meV superconducting gap, a spin excitation, or some type of interaction with the theorized stripe phase.  

The experimental system of the current work measures the very small complex Faraday angle imparted to CO$_2$ laser radiation traveling perpendicular to and transmitted by the sample immersed in a perpendicular magnetic field.  The system is the same as that used by in the earlier YBCO study~\cite{CernePRL2000} with the addition of an inline calibration system and a continuous stress-free temperature scan provision, and further modified to place a difraction-limited spot on the sample with post-sample spatial filtering~\cite{Cerne2002}.

The sample of the current work was cleaved or rather, peeled from a single bulk crystal of Bi$_2$Sr$_2$CaCu$_2$O$_{8+\delta}$ grown by the floating zone method and then placed in thermal contact with a wedged BaF$_2$ crystal.  Measurement of the AC magnetic susceptance of this mounted peeled segment revealed a T$_{\text{c}}$ of 92 K with a width of less than 1K.  This measurement, performed after all of the Hall measurements had been completed, establish the integrity of the sample and recommend the Hall data as representative of optimally doped BSCCO.  Infrared conductivity data from measurements performed on bulk crystals from the same batch supplied the real and imaginary parts of $\sigma_{xx}$ required to obtain the Hall angle from the Faraday angle data.  The calculation of $\sigma_{xy}$ considered the multilayer reflection effects within the BSCCO film~\cite{Cerne2002}.

To form $\theta_{\text{H}}$\ one divides $\sigma_{xy}$ by $\sigma_{xx}$ corresponding to the same temperature and frequency.  Since we wish to examine that part of the Hall angle related to the free carriers it is necessary to remove the contributions to $\sigma_{xx}$ coming from interband transitions which occur above $\sim$1 volt.  These transitions contribute essentially nothing to $\sigma_{xy}$ at the frequencies of these experiments.  Therefore, before taking the ratio, we removed their contribution to the imaginary part of $\sigma_{xx}$ using the value $\epsilon_{\text{infinity}} = 4.6$ from Quijada \cite{QuijadaThesis}. This increases Re($\theta_{\text{H}}$) by about 5\% and reduces Im($\theta_{\text{H}}$) by about 5\%.

\begin{figure}
\includegraphics[width=8.6cm, clip=true]{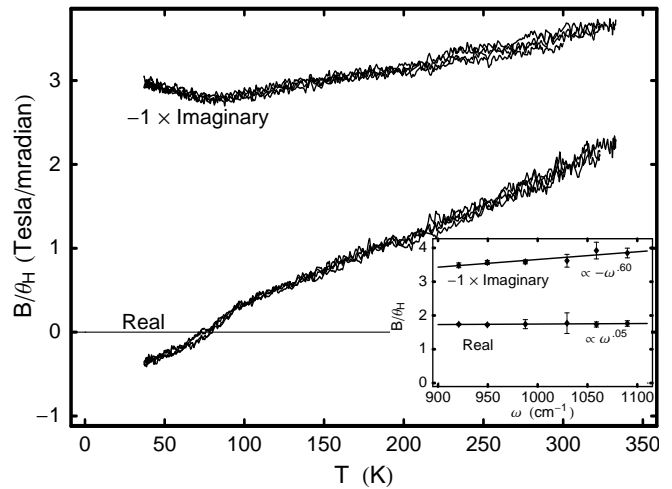}
\caption{\label{fig;inverseThetaH} The real and imaginary parts of $\theta_{\text{H}}^{-1}$ for 2212 BSCCO versus temperature.  The inset displays $\theta_{\text{H}}^{-1}$ at 300K and versus wavenumber .}
\end{figure}

\begin{figure}
\includegraphics[width=8.6cm, clip=true]{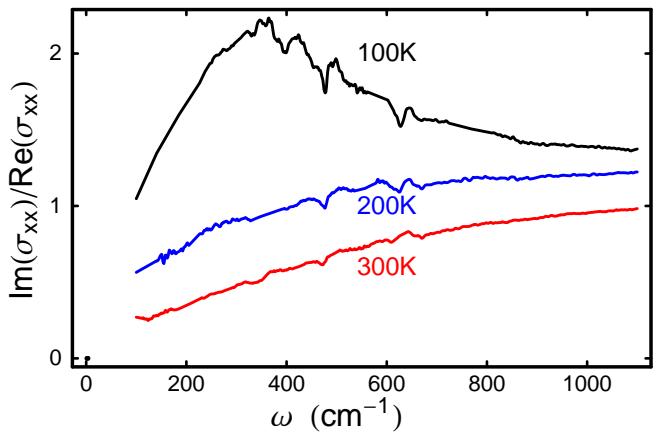}
\caption{\label{fig;phaseSigmaVersusF}$\text{Im}\left(\sigma_{xx}\right)/\text{Re}\left(\sigma_{xx}\right)$ vs. frequency for single crystal 2212 BSCCO at different temperatures.}
\end{figure}

Figure~\ref{fig;inverseThetaH} displays the resulting values of $\theta_{\text{H}}^{-1}$.  The most striking feature of the mid-infrared Hall angle response is its strong temperature dependence in comparison with the temperature dependence of $\sigma_{xx}$ at these frequencies~\cite{TuPRB2002}. The observed response is not consistent with the Lorentzian squared form found in YBCO at far-infrated frequencies which, in particular, implies a positive Re$(\theta_{\text{H}})$ at these frequencies. Indeed the response in the mid-infrared is found to be close to that of a Drude model $\theta_{\text{H}}=\omega_{H}/(\gamma-i\omega)$. The Re$(\theta_{\text{H}}^{-1})$ in Fig.~\ref{fig;inverseThetaH} is nearly frequency independent but strongly temperature dependent. The Im$(\theta_{\text{H}}^{-1})$ is only weakly temperature dependent and has a power law frequency dependence at 300 K but is of somewhat lower degree than the linear dependence $\omega$ expected for a Drude model. This suggests that by $1000\ \text{cm}^{-1}$ the system is approaching but has not yet reached the high frequency asymptotic Drude form. The condition for achieving this asymptotic form are that the frequency must be higher than all the characteristic frequencies of the system, including the quasiparticle relaxation rate.  While 100 meV is higher than the superconducting gap, the phonon energies and the neutron magnetic resonance it is still below the estimated antiferromagnetic coupling energy $J\approx200 \text{meV}$ for the cuprates. Indeed the longitudinal conductivity has not achieved the widely reported  asymptotic behavior as can be seen in Fig.~\ref{fig;phaseSigmaVersusF}, where by 1000 cm$^{-1}$ $\text{Im}(\sigma)/\text{Re}(\sigma)$ has become nearly frequency independent in contrast to the linear frequency dependence for a Drude response.  

\begin{figure}
\includegraphics[width=8.6cm, clip=true]{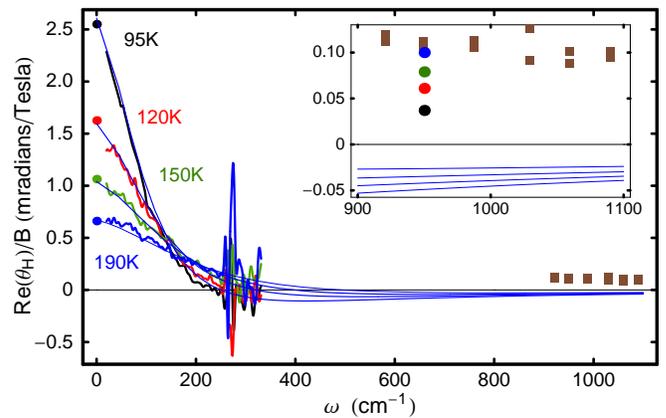}
\caption{\label{fig;thetaHVersusFwfar} The real and imaginary parts of $\theta_{\text{H}}/Tesla$ versus temperature for 2212 BSCCO.  The mid-infrared data is shown in the inset with expanded axes.  The brown points in the inset correspond to 300 K.  The other colors of the points at  950 cm$^{-1}$ within the inset corresond to the same temperatures as in the main plot.}
\end{figure}

Figure~\ref{fig;thetaHVersusFwfar} compares the real part of the Hall angle with that from the far-infrared data~\cite{GraysonPRL2002}.  We observe a disconnect between the behavior of the far- and mid-infrared Hall angle.  Starting at about 230 cm$^{-1}$ the real part of the Hall angle for low temperatures appears to be negative.  However, as noted above it is positive at all temperatures above T$_{\text{c}}$ in the mid-infrared.  The thin lines are fits of the full complex far-infrared Hall angle data to squared Lorentzians.  However, the extension of the real part into the mid-infrared remains negative unlike the data which are positive.  There are numerous ways to accommodate this  plunge in the real part of the Hall angle at 300 cm$^{-1}$ and to join these results with those of the mid-infrared of the current work.  All would require at least one minimum and at least one maximum between 300 and 1000 cm$^{-1}$ suggesting a resonance in this frequency interval.  No such feature is seen in the data for $\sigma_{xx}$~\cite{TuPRB2002}.  However, a peak is present at ~300 cm$^{-1}$ in the phase of $\sigma_{xx}$ in the plot of $\mbox{Im}\left(\sigma_{xx}\right)/\text{Re}\left(\sigma_{xx}\right)$ at 100 K shown in Fig.~\ref{fig;phaseSigmaVersusF}.  It is interesting to note that this feature in the phase is not predicted by most models that have been proposed for the conductivity in the cuprates, for example, the coldspot  model~\cite{Ioffe1998}.  Further, its frequency roughly corresponds to numerous other phenomena including the 41 meV spin resonance, the 50$\pm$15 meV phonon interaction~\cite{Bogdanov2000}, the superconducting gap, and the quasiparticle band width of the three band Hubbard model.

\begin{figure}
\includegraphics[width=8.6cm, clip=true]{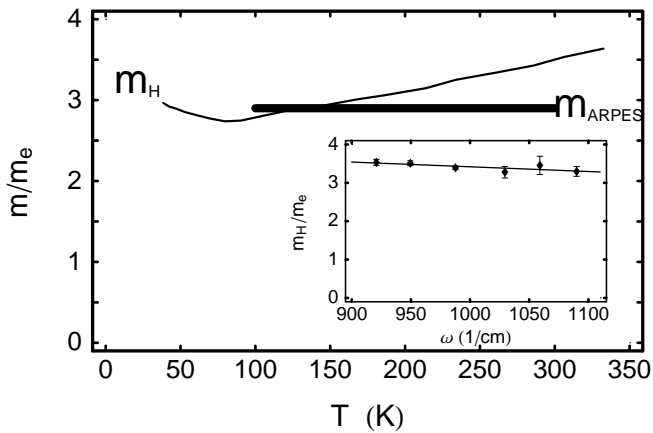}
\caption{\label{fig;massVersusT}The mid-infrared Hall mass (thin line) and the ARPES Fermi surface mass (thick line) both versus temperature.  The mid-infrared Hall mass is also shown in the inset versus frequency.}
\end{figure}

\begin{figure}
\includegraphics[width=8.6cm, clip=true]{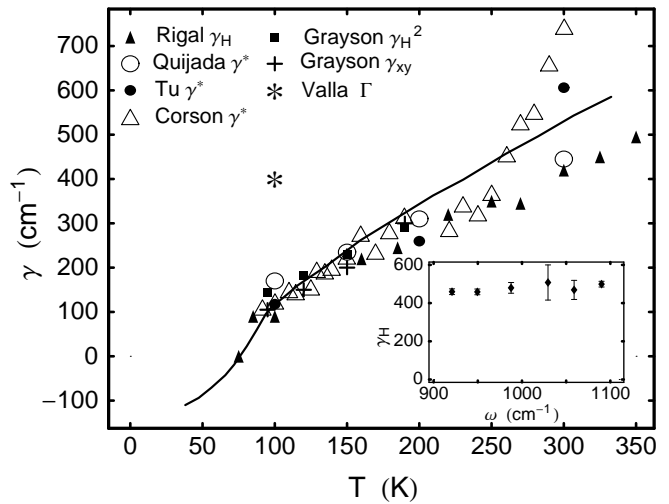}
\caption{\label{fig;gammaVersusT} The scattering rates from different measurements compared.   The thick black line demonstrates the temperature dependence of the Hall scattering rate in the mid-infrared region.  The mid-infrared Hall scattering rate is also shown in the inset versus frequency.}
\end{figure}

In view of the near Drude behavior of the mid-IR Hall data we have used the Drude model to extract the parameters of the model $\omega_{\text{H}}$ and $\gamma_{\text{H}}$ and to examine their frequency and temperature dependence. We express $\omega_{\text{H}}$ in terms of the Hall mass defined as $m_{\text{H}}=eB/\omega_{\text{H}}c$. The data are shown in Figs.~\ref{fig;massVersusT} and~\ref{fig;gammaVersusT}. The Hall mass is seen to be in remarkably good agreement, especially at $T_{c}$, with the hole effective mass at the Fermi level deduced from ARPES measurements. The inset data in Fig.~\ref{fig;massVersusT} indicates a decreasing mass with frequency in accord with the general expectation that it approach the band value estimated at about half the value at $E_{\text{F}}$. The data also indicate an increasing mass with temperature, which is quite counter to the expectations for strongly interacting systems. However, we note that $\gamma_{\text{H}}$ and the carrier relaxation rate in general is an increasing function of $T$ achieving its smallest value at $T_{c}$. As the Hall angle (or inverse Hall angle) can be expanded asymptotically as a series in $z=\gamma_{n}/\omega$ (where $\gamma_{n}$ are various averages of the scattering rates) the asymptotic behavior is most closely achieved at the lowest temperatures where z is small. Indeed it is seen that $m_{\text{H}}$ is in excellent agreement with ARPES at low $T$ and deviates as $T$ is increased.

Figure~\ref{fig;massVersusT} displays a Hall mass, which is roughly twice the value of the longitudinal mass calculated from $\sigma_{xx}$ in the mid-infrared.  The frequency dependences for the Hall mass and longitudinal mass are similar displaying a slow decrease with increasing frequency.  However, the temperature dependences of the masses are opposite each other with the Hall mass actually increasing about 20\% over the temperature range of 100 to 300K.  Finally, the same ARPES measurements which imply a very different scattering rate find a full dressed mass at the Fermi level which is nearly equal to the measured mid-infrared Hall mass.

The linear temperature dependence of the Hall scattering rate in the mid-infrared also disagrees with the ARPES results for $\text{Im}\Sigma$~\cite{VallaScience1999}, which imply a decreasing temperature dependence with increasing frequency as is also seen in $\sigma_{xx}$.  Also marginal Fermi liquid theory, which posits a linear temperature dependence of the scattering is at odds with the lack of frequency dependence.  Any significant frequency dependence would cause a significant positive intercept in the zero temperature projection of the normal state scattering rate in Fig.~\ref{fig;gammaVersusT}.  The lack of frequency dependence of the scattering rate of the current work on the other hand contravenes both Fermi liquid and marginal Fermi liquid theories, which claim a strong frequency dependence for the scattering rate at temperatures low compared to the photon energies of the experiment.  We find little or no frequency dependence of the scattering rate at $\omega=1000\approx 125\, \text{meV}$.  The far-infrared scattering rate when compared to that of the mid-infrared exhibits the same linear increase with temperature.  However, the projection of the normal state  mid-infrared scattering rate to zero temperature is negative.  This feature was also observed for YBCO~\cite{CernePRL2000}.

Figure~\ref{fig;gammaVersusT} also shows the relaxation rates deduced from measured quantities from the mid-IR Hall data and a number of other experiments.  These include IR conductivity $\gamma^*$ for BSCCO~\cite{TuPRB2002} determined from the slope in Fig.~\ref{fig;phaseSigmaVersusF} extrapolated to zero frequency :
\begin{equation}
\gamma^{*}=\frac{\gamma}{1+\lambda(\omega)}=-\omega \, \frac{\text{Re}(1/\sigma)}{\text{Im}(1/\sigma)}=\omega \, \frac{\text{Re}(\sigma)}{\text{Im}(\sigma)}
\end{equation}
Also included are: the relaxation rate deduced from $\sigma_{xx}$ for YBCO extrapolated to zero frequency designated ``Quijada $\gamma^*$''~\cite{QuijadaThesis}; the Hall relaxation rates designated ``Grayson $\gamma_{\text{H}}^2$'' and ``Grayson $\gamma_{xy}$'' both for YBCO in the far-infrared, the first using using the squared Lorentzian analysis and the second using a simple Drude applied to $\sigma_{xy}$~\cite{GraysonPRL2002}; the terahertz relaxation rate designated ``Corson $\gamma^*$''~\cite{Corson2002} deduced from $\gamma^*$ for BSCCO at  0.2-1.0 THz; the Hall relaxation rate for YBCO designated ``Rigal $\gamma_{\text{H}}$'' deduced from $\theta_{\text{H}}$~\cite{RigalData}; and the ARPES zero frequency value for $\Gamma$ for BSCCO designated ``Valla $\Gamma$''~\cite{VallaPRL2000}. Except for the ARPES result, the different relaxation rates are in remarkable agreement with each other, especially at low $T$. However, they are all significantly smaller than the hole scattering rates at $E_{\text{F}}$ deduced from ARPES in the $(\pi, \pi)$ direction where the linewidths take on their minimum values.

These results suggest a universal linear $T$ relaxation rate in the hole doped cuprates. More remarkably they also suggest that the transport relaxation rate is temperature dependent but not frequency dependent. In this case the apparent frequency dependence of the infrared relaxation rate deduced from $\sigma$ may not represent an inelastic scattering effect but drives instead from some other mechanism such as that suggested by Ioffe and Millis. Therefore these results suggest that current relaxation in the optimally doped cuprates is quasi-elastic.     

We acknowledge fruitful discussions with C. Varma, E. Abrahams, V. Yakovenko, and A. Millis.  This work was supported by the NSF under grant DMR-0070949 and by DOE under contract no. DE-AC02-98CH10886. 

\bibliography{BSCCOMidHall}

\end{document}